\documentstyle[epsf,twocolumn]{esa_sp_latex}

\def\a218{\alpha_{2-18}}

\def\aint218{\alpha_{{\rm int},2-18}}

\newcommand{\bez}{\begin{eqnarray*}}
\newcommand{\eez}{\end{eqnarray*}}
\newcommand{\be}{\begin{equation}}
\newcommand{\ee}{\end{equation}}

\newcommand{\beq}{\begin{eqnarray}}
\newcommand{\eeq}{\end{eqnarray}}
\newcommand{\bc}{\begin{center}}
\newcommand{\ec}{\end{center}}

\newbox\grsign \setbox\grsign=\hbox{$>$} \newdimen\grdimen \grdimen=\ht\grsign
\newbox\simlessbox \newbox\simgreatbox \newbox\simpropbox
\setbox\simgreatbox=\hbox{\raise.5ex\hbox{$>$}\llap
     {\lower.5ex\hbox{$\sim$}}}\ht1=\grdimen\dp1=0pt
\setbox\simlessbox=\hbox{\raise.5ex\hbox{$<$}\llap
     {\lower.5ex\hbox{$\sim$}}}\ht2=\grdimen\dp2=0pt
\setbox\simpropbox=\hbox{\raise.5ex\hbox{$\propto$}\llap
     {\lower.5ex\hbox{$\sim$}}}\ht2=\grdimen\dp2=0pt

\begin{document}
% esa_sp_latex.tex
% PLEASE DO NOT EDIT THIS FILE. THANK YOU
%

\parindent 0pt
\parskip 10pt plus 1pt minus 1pt
\hoffset=-1.5truecm
\topmargin=-1.0cm
\textwidth 17.1truecm \columnsep 1truecm \columnseprule 0pt % FF

\topmargin 1cm

\title{\bf AVERAGE PROPERTIES OF THE TIME BEHAVIOUR OF GAMMA-RAY BURSTS} 

\author{{\bf Boris~Stern$^{1,2}$, Roland~Svensson$^2$
and Juri~Poutanen$^{2,3}$} \vspace{2mm} \\ 
$^1$Institute for Nuclear Research, Moscow \vspace{2mm} \\
$^2$Stockholm Observatory, Saltsj\"obaden, Sweden \vspace{2mm} \\
$^3$Uppsala Observatory, Uppsala, Sweden} 

\maketitle

\begin{abstract}
Time profiles of gamma ray bursts (GRBs) are extremely diverse in their
durations, morphologies, and complexity. Nevertheless, the average peak-aligned
profile of all bursts detected by BATSE with sufficient data quality
has a simple ``stretched'' exponential shape,
$F \propto \exp[-(t/t_0)^{1/3}]$, where $t$ is the
time measured from the time for the peak flux, $F_p$, of the
event, and $t_0$ is a time constant. 
We study the behaviour of $t_0$ of both the post-peak and the pre-peak slopes of 
the average time profile as a function of the  peak brightness range of the burst sample.
We found that the post-peak slope shows time 
dilation when comparing bright and dim bursts, while the pre-peak slope hardly  
changes. Thus dimmer bursts have a different shape -- they are more asymmetric.
This shape-brightness correlation is observed at a 99.6\% confidence
level. Such a correlation has a natural explanation within the
pulse avalanche model, which is briefly described. Complex events, 
consisting of many pulses are more symmetric and are intrinsically brighter. 
Bursts consisting of one or a few pulses are intrinsically 
weaker and more asymmetric. For such a correlation to be observable requires
that the luminosity distance distribution of GRBs to be  different from a power-law.


\end{abstract}

\section{INTRODUCTION}

Stern (1996) found that the average peak-aligned time profile of GRBs
(the procedure of peak-alignment was pioneered by Mitrofanov et al. 
1994, 1995) in the BATSE-2 catalog
has a simple ``stretched'' exponential shape, \\
$<F/F_p>=\exp[-(t/t_0)^{1/3}]$, where $t$ is the
time since the peak flux, $F_p$, of the
event, and $t_0$ is a constant ranging from 0.3 s for strong bursts
to $\sim$ 1 s for dim bursts. This dependence of $t_0$ on brightness
can be interpreted as a  cosmological time dilation (e.g., \cite{pac92,pir92}).

Such a simple average time profile is remarkable considering the diverse and
chaotic behavior of the individual time profiles of GRBs.
On the other hand, the simple shape of the average time profile gives an excellent opportunity
to study effects such as time dilation.

Here, we study the two slopes of the average time profile for a larger sample of GRBs  
and with a more accurate treatment of the background than was done in 
Stern (1996). Another advantage we now have is the access to the
pulse avalanche model developed by Stern \& Svensson (1996) which successfully 
describes many 
statistical properties of GRBs including the stretched exponential shape 
of the profile and, of particular importance for the present work, the 
rms variance of individual time profiles. This means that we can rely on
this model when estimating the errors of stretched exponential fits, which  in 
turn gives us reliable estimates of the significance 
levels of the observed effects.

\section{DATA PROCESSING} 

This work is based on data obtained from the publicly available BATSE 
database at Goddard Space Flight Center. Our sample includes bursts
up to trigger number 3745. We used the 0.064 s and 1.024 s time resolution
data from the Large Area Detectors (LAD). All time profiles were constructed 
with 64 ms time resolution together with pre and post-burst extensions
of 1024 ms time resolution. All 
background fits were done with the 1024 ms data as they cover a wider time interval
including the  pre-trigger history. 
The time profiles were studied using count rates in all 4 LAD's energy channels.

The  procedure of background fitting included: 

-- A visual examination of all bursts including both 64ms and 1 s resolution data.

-- All doubtful peaks and count rate variations with
$\chi^2$ exceeding that of Poisson noise were analyzed in order to see
whether they came from the same direction as the main peak of the burst.
This was accomplished by comparing the direction of the eight-component
vector consisting of the $\chi^2$ from the eight LADs for the feature
with the direction of the corresponding vector for the main peak.

-- Discarding all events with highly variable background as well as those where we were 
unable to confidently extract  sufficiently long intervals of pre-peak and 
post-peak histories.

-- Possible use of widely separated  
fitting windows to avoid losses of weak  GRB signals.

When sorting bursts into brightness groups we used peak fluxes for 64 ms 
time resolution from the BATSE-3 catalog.

The errors were calculated using the pulse avalanche model. What we need are the
statistical errors of the stretched exponential fits to the data. To extract
this error directly from the data is very difficult because of strong 
correlations along the time profile. Instead,  we simulated many samples 
of $N$ events, determined the average time profile for each sample, and then 
made a stretched exponential fit to each simulated average profile:
$F(t)=\beta \exp[-(t/t_0)^{1/3}]$, 
where $t_0$ and $\beta$ are fitting parameters.
The same procedure was used as for real data.  Finally,
the rms errors of the fitting parameters was calculated.

We have two slopes of the profile - the pre-peak (rising) slope 
and post-peak (decaying) slope. 
We fitted them simultaneously with different $t_0$, denoted here as 
$t_r$ and $t_d$, and with a common $\beta$.
Fitting the simulated samples gave the following relative standard deviations:  

$\sigma (t_{r,d})/ t_{r,d} = 0.20 \sqrt{100/N}$ 

$\sigma (t_r + t_d) / (t_r + t_d) = 0.19 \sqrt{100/N}$

$\sigma (t_d/t_r) / (t_d/t_r) = 0.13 \sqrt{100/N}$

Note, that the relative accuracy for the sum of the two time constants is close to that
for one time constant, while the accuracy of the asymmetry ratio,
$t_d/t_r$, is markedly better.
This is a consequence of a strong correlation between the two slopes -- 
a circumstance that 
favors the measurement of shape vs. brightness correlations and that complicates 
the measurement of the time dilation effect.
The errors are robust against variations of the parameters of the pulse
avalanche model as long as  
the model gives approximately the correct stretched exponential average profile.
The procedure for the  data analysis will be described in greater detail in 
Stern et al. (1997).

\section{THE AVERAGE TIME PROFILE FOR DIFFERENT BRIGHTNESS GROUPS} 

The average peak-aligned profiles for three  brightness groups are shown
in Figure 1. Both the rising and and the decaying profiles are well shaped 
stretched exponentials
for the bright and the medium group, while both profiles are quite  deformed for the 
weakest group. The rising (pre-peak) slope is steeper for all brightness groups,
but the asymmetry is increasing when going from the brightest to the dimmest group.
The results of our stretched exponential fits are summarized in Table 1
and in Figure 2.

\begin{figure}[htbp]
  \begin{center}
    \leavevmode
\epsfxsize=8.0cm  
\epsfbox[74 340 460 740]{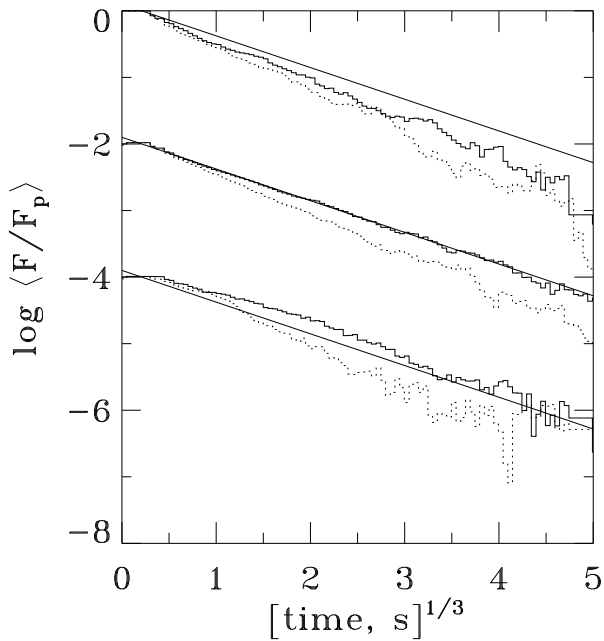}
  \end{center}
  \caption{\em 
Average peak-aligned time profiles for three brightness groups:
1.) $ F_p >$ 5  ph cm$^{-2}$ s$^{-1}$, 157 GRBs (upper curves); 
2.) 0.7 $< F_p <$ 5, 630 GRBs (middle curves); and 
3.)  $ F_p <$ 0.7, 116 GRBs (lower curves). Middle and lower curves 
are shifted downwards for clarity. 
Solid and dotted curves represent the average post-peak and pre-peak 
time profiles, respectively. 
Straight lines show the best linear fit to the post-peak history
of the medium brightness group. }
 \label{fig:1} 
\end{figure}

\begin{figure}[htbp]
  \begin{center}
    \leavevmode
\epsfxsize=8.0cm
\epsfbox[60 380 460 720]{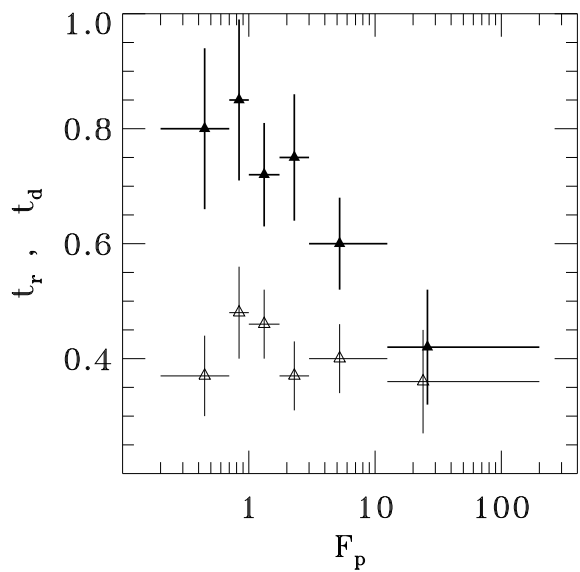}
  \end{center}
  \caption{\em
Time constants, $t_r$ and $t_d$,  vs. peak photon flux, $F_p$, 
in 64 ms time resolution. 
Lower and upper crosses represent $t_r$ and $t_d$
for the pre-peak (rising)  and post-peak (decaying) 
average time profile, respectively. Error bars of the time constants correspond
to 1$\sigma$. Error bars in photon flux represent 
the width of the brightness groups. }
 \label{fig:2}
\end{figure}

\begin{table*}[tn]
\begin{center}
\begin{tabular}{l l l l l l l} 
\hline
%\tableline
\# & Peak flux &  $N$ &  $t_d$ & $t_r$ &  $t_r$ + $t_d$ &  $t_d/t_r$ \\
\hline
%\tableline
1 & 12.5 -- 200  & 64  & 0.42$\pm$ 0.10 & 0.36$\pm$0.09 & 0.78 $\pm$ 0.18 & 1.17 $\pm$0.19 \\
2 & 3 -- 12.5 & 193 & 0.60$\pm$ 0.08 & 0.40$\pm$0.06 & 1.00 $\pm$ 0.13 & 1.50 $\pm$0.11 \\
3 & 1.75 -- 3 & 159 & 0.75$\pm$ 0.11 & 0.37$\pm$0.06 & 1.12 $\pm$ 0.16 & 2.02 $\pm$0.21 \\
4 & 1 -- 1.75 &  241 & 0.72$\pm$ 0.09 & 0.46$\pm$0.06 & 1.18 $\pm$ 0.14 & 1.56 $\pm$0.13 \\
5 & .7 -- 1 & 139 & 0.85$\pm$ 0.14 & 0.48$\pm$0.08 & 1.33 $\pm$ 0.21 & 1.89 $\pm$0.18 \\
6 & 0 -- .7  & 116 & 0.80$\pm$ 0.14 & 0.37$\pm$0.07 & 1.17 $\pm$ 0.20 & 2.16 $\pm$0.26 \\
7 & 7.5 -- 200 & 111 & 0.47$\pm$ 0.08 & 0.36$\pm$0.07 & 0.83$\pm$ 0.14 & 1.30 $\pm$0.16 \\ 
8 & 5 -- 200  & 157 & 0.50$\pm$ 0.07 & 0.35$\pm$0.06 & 0.85 $\pm$ 0.12 & 1.56 $\pm$0.13 \\
9 & .75 -- 2.5   & 463 & 0.78$\pm$ 0.07 & 0.45$\pm$0.06 & 1.22 $\pm$ 0.11 & 1.73 $\pm$0.10 \\
\hline
%\tableline
\end{tabular}
\end{center}
\caption{Time constants [s] 
of the stretched exponential fit to the averaged 
pre-peak ($t_r$) and post-peak ($t_d$) profiles. Peak flux 
[ph cm$^{-2}$ s$^{-1}$] is taken from BATSE database. 
$N$ is the number of bursts in the given brightness interval. }
%\tablecomments{(to remove $\chi^2$!)}
\end{table*}

Fits of all time profiles have a good $\chi^2$ except that for the weakest sample (\# 6).
Its profile is apparently deformed. This deformation is partially associated 
with the low trigger efficiency of this sample and with effects of Poisson noise.
Nevertheless these biases are insufficient to explain this deformation and
a real deformation effect could be present in the shape of the weakest sample. 
An analysis of this effect is beyond the scope of this poster paper. 

The strongest effect in the remaining samples is the time dilation of the decaying 
(post-peak) slope of the
time profile. Comparing $t_d$ for samples 1 and 9 we find the time dilation to be a
factor 1.86$^{+0.82}_{-0.57}$, where the 90\% confidence interval is given. 
Comparing samples 7 and 9 gives a factor  1.66$^{+0.68}_{-0.42}$.
Varying the lower flux limit of 
the brightest sample from $ F_p$ = 5 ph cm$^{-2}$ s$^{-1}$ and upwards
does not significantly change the magnitude of the time dilation effect.
A careful estimate of the significance level of the time dilation using 
model simulations gives 0.985.
 
The rising (pre-peak) slope is, however, surprisingly stable. 
The variations of $t_r$ do not exceed 
statistical errors. This leads to an increasing asymmetry ratio, $t_d/t_r$, for the
weaker samples. The effect is significant  comparing samples 1 and 9: 
$(t_d/t_r)_{\rm dim}/(t_d/t_r)_{\rm bright}$ = 1.48$^{+0.55}_{-0.32}$ 
($90\%$ confidence interval). The probability of sampling such
ratios if the asymmetry was constant is 4 $\cdot$ 10$^{-3}$.  

The time dilation of $t_r + t_d$ has a smaller, but still acceptable significance
when comparing samples 7 and 9. The ``dim/bright'' ratio becomes 1.47$^{+0.63}_{-0.36}$.
The probability for zero time dilation is 0.01.

\section{INTERPRETATION OF CORRELATIONS USING THE PULSE AVALANCHE MODEL} 
 
 The pulse avalanche model is based on the assumption that the well-shaped 
stretched exponential time profile results from a simple stochastic
 process responsible for the generation of time histories
of GRBs. Then the diversity of  GRB's time histories arise as different
random realizations of the same stochastic process at approximately the same 
parameters. The important requirement is a near-criticality of the process - 
then it provides a large variety of individual bursts behaviours.

 This idea was implemented as a near-critical chain reaction of events
(of still unknown nature, it could, e.g., be reconnections of 
turbulent magnetic field), where each event is associated with one pulse of 
gamma-ray emission. Then, in a near critical regime, one spontaneous pulse can 
give rise to a long cascade of secondary pulses piling up into a complex
chaotic event, or, depending on chance, no further pulses may result
and we will instead see a simple single pulse event.

With a proper scaling of time delays between pulses and a proper spectrum of 
spontaneous pulses (flicker noise), the model successfully reproduces  the 
stretched exponential profile (see Figure 3) as well as the autocorrelation 
function. There is also qualitative  agreement with the duration distribution,
the ratio of the number of simple and
complex bursts is reproduced, and the model even produces the visual
impression of real bursts. For more details see Stern \& Svensson (1996).

Just two simple assumptions  
of those implemented in the pulse avalanche model 
are needed to demonstrate that all correlations 
described above are very natural. GRBs consist of a number of pulses of 
different durations but similar shapes. Let all these pulses have locally
independent sources of energy. Then if two pulses coincide in time, their 
amplitudes sum up. In a complex event, hundreds of pulses are piling up, 
increasing the peak brightness by up to an order of magnitude. 
Then, if a pulse is a kind of
standard candle, simple events are intrinsically weak and complex events
are intrinsically bright.

At the same time, simple events are asymmetric just because
a single pulse is asymmetric (e.g., Norris et al. 1996)
with a sharp rise and a slower decay. 
This asymmetry
is washed out in complex events where the position of the highest peak is 
more or less random among many overlapping pulses. 

To demonstrate this we simulated a large sample of ``bursts'' using model 
parameters which gave an  approximate  agreement with the stretched 
exponential average time profile for the sample of all real bursts.
The results are summarized in Table 2.
We see that the asymmetry ratio, $t_d/t_r$, increases with decreasing peak flux.

\begin{table}[h]
\begin{center}
\begin{tabular}{l l l l l l} 
%\tableline
\hline
Peak flux &  &  $t_d$ & $t_r$ &  $t_r$ + $t_d$ &  $t_d/t_r$ \\
%\tableline
\hline
0 -- $\infty$   &  & 0.66 & 0.41 &  1.07 & 1.61  \\ 
3 -- $\infty$   &  & 0.75 & 0.68 &  1.43 & 1.10 \\ 
0.8 -- 3        &  & 0.72 & 0.50 &  1.22 & 1.44  \\
0 -- 0.8        &  & 0.50 & 0.19 &  0.69 & 2.63 \\
\hline 
%\tableline
\end{tabular}
\end{center}
\caption{Time constants of the simulated bursts in different ``intrinsic
brightness'' intervals. The amplitude of the single pulse is sampled 
uniformly from the interval $[0,1]$.}
\end{table}

\begin{figure}[htbp] 
  \begin{center}
    \leavevmode
\epsfxsize=7.9cm  
\epsfbox[70 380 460 685]{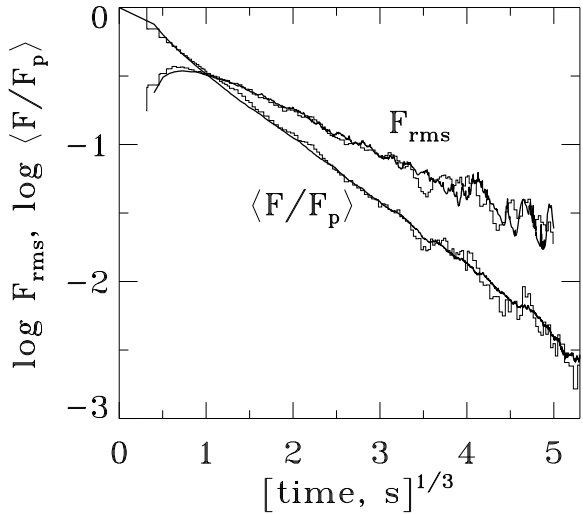}
  \end{center}
  \caption{\em 
Comparison between observed and simulated average post-peak
time profiles of GRBs. 
Thin-line histogram labelled $<F/F_p>$ shows the average 
peak-aligned post-peak 
time profile for the 598 useful BATSE-3 events 
as the fractional flux,
$<F/F_p>$ vs. $t^{1/3}$, where $F_p$ is the peak flux and
$t$ = 0 -- 150 s is the time since the strongest peak.
Thick curve labelled $<F/F_p>$ is the average peak-aligned time profile
for 5000 simulated time profiles. The set of curves labelled
$F_{\rm rms}$ are the
rms deviations of individual peak-aligned time profiles,
$F_{\rm rms} \equiv [<(F/F_p)^2>-<F/F_p>^2 ]^{1/2}$,
for
both real (thin line histogram, 598 events) and
simulated (thick curve, 5000 events) time profiles. 
Further details are given by Stern (1996) and Stern \& Svensson (1996). }
 \label{fig:3} 
\end{figure}

\section{CONCLUSIONS} 

Besides the time dilation effect, observed in many previous works 
(e.g., Norris et al. 1994) 
we also see a dependence of the profile asymmetry on brightness and this effect 
is of the same order of magnitude as the time dilation itself.

A correlation of such a kind cannot be due to spectral redshifts.
We find that 
strong events have slightly smaller asymmetry in the higher energy bands,
(LAD channels 4 and 3) than in the lower energy bands (LAD channels 1 and 2).
Redshifting the softer part of the spectrum below the detector threshold
would then give rise to more symmetric, rather than asymmetric profiles.
Details will be published elsewhere.

The effect of trigger efficiency is negligible for the brightness groups
considered here. Also one can hardly  find an evolutionary factor that would
change the asymmetry.
One can suggest that there exist two separate classes of GRBs with
different degree of asymmetry, which are differently distributed in space
(separate classes of {\it long} bursts are required as short bursts do
not contribute to the asymmetry).
But such a suggestion seems too arbitrary, too radical, and unnecessary
as there exists a much simpler explanation.

The simplest explanation is that the observed correlation is a consequence of
an intrinsic correlation between shape and brightness as described above.
The necessary condition for such a correlation to be observable is a
significant deviation from a power law for the GRB distribution over luminosity
distance. This would allow intrinsically strong events to dominate in
the brightest observational range. The observed log $N$ - log $P$ distribution
is actually curved (Meegan et al. 1996) 
and this is natural if the distance distribution
covers both Euclidean and $z \sim 1$ regions which have different
luminosity distance scalings.

Maybe our detection of a strong shape - brightness 
 correlation imposes a stronger constraint
on the curvature of the true radial distribution of GRBs than what follows from
the observed log $N$ - log $P$ distribution. Detailed studies are, however, required
 to formulate this intuitive conclusion at a quantitative level.

As one kind of correlation has been observed, other kinds of intrinsic
correlations may also be observable and this causes a problem for the cosmological
interpretation of the time dilation effect. This problem, considered by
Brainerd (1994), arises from unavoidable correlations between peak luminosity
and time scales caused by  different bulk Lorentz factor in the sources of
different bursts. This effect can mimic both cosmological time dilation and
spectral redshift.

However, if our interpretation using the pulse avalanche model is valid we must
conclude that the real time dilation is larger than that obtained from Table 1.
Actually, within the pulse avalanche framework, intrinsically weak events are 
not only more asymmetric, but they are also narrower (see column $t_r+t_d$
in Table 2).
This is a new correction that increases the real time dilation more
than the correction arising from  spectral redshift (see Norris et al. 1994).
The corrected time dilation could exceed a factor 2 and it could be caused 
by different effects, including the cosmological one. Unfortunately, the task
of extracting the cosmological component from the total time dilation
seems extremely difficult. 

\section*{ACKNOWLEDGMENTS}

We acknowledge support from the Swedish Royal Academy of Sciences, 
the Swedish Natural Science Research Council 
(in particular a postdoctoral fellowship for J.P.), 
and a NORDITA Nordic Project grant. 
This research used data obtained 
from the HEASARC Online Service provided by NASA/GSFC.


\begin{thebibliography}{}

\bibitem[\protect\astroncite{Brainerd}{1994}]{br94}
Brainerd, J.~J. 1994, ApJ, 428, L1 

\bibitem[\protect\astroncite{Meegan et al.}{1996}]{meg96}
 Meegan, C.~A., et al. 1996, ApJS, 106, 65

\bibitem[\protect\astroncite{Mitrofanov et al.}{1994}]{mit94}
Mitrofanov, I. G., Chernenko, A. M., Pozanenko, A. S., Paciesas, W. S.,
Kouveliotou, C., Meegan, C. A.,  Fishman, G. J.,  Sagdeev, R. Z. 
1994,
\newblock in G. J. Fishman, J. J. Brainerd, K. C. Hurley (eds), 
Proc. of the Second Huntsville workshop on Gamma Ray Bursts,
AIP, New York, 187  

\bibitem[\protect\astroncite{Mitrofanov et al.}{1995}]{mit95}
Mitrofanov, I. G., et al.
1995,
Astronomy Reports, 39, 305

\bibitem[\protect\astroncite{Norris et al.}{1994}]{nor94}
Norris, J. P., Nemiroff, R. J., Scargle, J. D., Kouveliotou, C., Fishman, G. J.,
Meegan, C. A., Paciesas, W. S.,  Bonnell, J. T.
1994,  ApJ, 424, 540

\bibitem[\protect\astroncite{Norris et al.}{1996}]{nor96}
Norris, J. P., Nemiroff, R. J.,  Bonnell, J. T., Scargle, J. D., 
Kouveliotou, C., Paciesas, W. S.,  Meegan, C. A.,  Fishman, G. J.
1996, ApJ, 459, 393

\bibitem[\protect\astroncite{Paczy\'nski}{1992}]{pac92} 
Paczy\'nski, B. 1992,  Nature, 355, 521

\bibitem[\protect\astroncite{Piran}{1992}]{pir92} 
Piran, T. 1992,  ApJ, 389, L45

\bibitem[\protect\astroncite{Stern}{1996}]{ste96}
Stern, B.~E. 1996, ApJ, 464, L111 

\bibitem[\protect\astroncite{Stern \& Svensson}{1996}]{sv96}
Stern, B.~E., Svensson, R. 1996, ApJ, 469, L109

\bibitem[\protect\astroncite{Stern et al.}{1997}]{s97}
Stern, B.~E., Poutanen, J., Svensson, R. 1997, in preparation 

\end{thebibliography}
\end{document}